\documentclass[fleqn,usenatbib]{mnras}
\usepackage{amsmath,amstext,mathrsfs}
\usepackage{txfonts}
\usepackage{float}
\usepackage{color} 
\usepackage{natbib}
\usepackage{subcaption}
\usepackage{gensymb}
\usepackage{natbib}
\usepackage{amssymb}
\usepackage{bm}
\usepackage{rotating}
\usepackage{graphicx,float}
\usepackage{caption}
\usepackage{tabularx}
\usepackage{longtable}
\usepackage{xcolor}
\usepackage{graphicx}
\usepackage{tikz}
\usepackage{soul}
\usepackage{footmisc}
\usepackage{comment}
\usepackage{makecell}
\usepackage{multirow}
\usepackage{ulem}
\usepackage[T1]{fontenc}

\usepackage{scalerel}
\usepackage{tikz}
\usetikzlibrary{svg.path}
\definecolor{orcidlogocol}{HTML}{A6CE39}
\tikzset{
	orcidlogo/.pic={
		\fill[orcidlogocol] svg{M256,128c0,70.7-57.3,128-128,128C57.3,256,0,198.7,0,128C0,57.3,57.3,0,128,0C198.7,0,256,57.3,256,128z};
		\fill[white] svg{M86.3,186.2H70.9V79.1h15.4v48.4V186.2z}
		svg{M108.9,79.1h41.6c39.6,0,57,28.3,57,53.6c0,27.5-21.5,53.6-56.8,53.6h-41.8V79.1z M124.3,172.4h24.5c34.9,0,42.9-26.5,42.9-39.7c0-21.5-13.7-39.7-43.7-39.7h-23.7V172.4z}
		svg{M88.7,56.8c0,5.5-4.5,10.1-10.1,10.1c-5.6,0-10.1-4.6-10.1-10.1c0-5.6,4.5-10.1,10.1-10.1C84.2,46.7,88.7,51.3,88.7,56.8z};
	}
}

\newcommand\orcidicon[1]{\href{https://orcid.org/#1}{\mbox{\scalerel*{
				\begin{tikzpicture}[yscale=-1,transform shape]
					\pic{orcidlogo};
				\end{tikzpicture}
			}{|}}}}

\DeclareRobustCommand{\VAN}[3]{#2}
\let\VANthebibliography\thebibliography
\def\thebibliography{\DeclareRobustCommand{\VAN}[3]{##3}\VANthebibliography}

\title[Study of Plasma turbulence]{Diagnosis of 3D magnetic field and modes composition in MHD turbulence with Y-parameter}

\author[Malik et al.]{
Sunil Malik
\orcidicon{0000-0003-4147-626X}$^{1,2}$
\thanks{E-mail:sunil.malik@uni-potsdam.de (SM)},
Ka Ho Yuen
\orcidicon{0000-0003-1683-9153}$^{3}$
\thanks{kyuen@lanl.gov (KHY)},
Huirong Yan
\orcidicon{0000-0003-2560-8066}$^{1,2}$
\thanks{huirong.yan@desy.de (HY) (Corresponding Author)}
\\
$^{1}$Institute fur Physik und Astronomie Universitat Potsdam, Golm Haus 28, D-14476 Potsdam, Germany\\
$^{2}$Deutsches Elektronen-Synchrotron DESY, Platanenallee 6, 15738 Zeuthen, Germany\\
$^{3}$Theoretical Division, Los Alamos National Laboratory, Los Alamos, NM 87545, USA
}

\date{Accepted XXX. Received YYY; in original form ZZZ}

\begin{document}
\label{firstpage}
\pagerange{\pageref{firstpage}--\pageref{lastpage}}
\maketitle

\begin{abstract}

Magnetic fields are crucial in numerous astrophysical processes within the interstellar medium. However, the detailed determination of magnetic field geometry is notoriously challenging. Based on the modern magnetohydrodynamic (MHD) turbulence theory, we introduce a novel statistical technique, the "Y-parameter", to decipher the magnetic field inclination in the ISM and identify dominant turbulence modes. The Y-parameter, calculated as the ratio of anisotropies of different Stokes parameter combinations, displays contrasting trends with the mean-field inclination angle in Alfv\'enic and compressible turbulence modes. A Y-parameter value around $1.5\pm0.5$ provide a statistical boundary to determine the dominant MHD turbulence modes. We have discovered specific correlations between the Y-parameter value and the inclination angle that unveil the dominant turbulence mode. This methodology, when applied to future radio polarisation surveys such as LOFAR and SKA, promises to significantly enhance our knowledge of 3D magnetic field in the ISM and improve our understanding of interstellar turbulence.

\end{abstract}

\section{Introduction}
\label{introduction}

The ISM is a complex, multi-phase environment comprising of gas, dust, and magnetic fields. Its magnetized and turbulent nature provides us with a good environment to study the magneto-hydrodynamic (MHD) turbulence in our galaxy~\citep{1995ApJ...443..209A,ElmegreenScalo,Boldyrev2006PhRvL,2007ARA&A..45..565M, Draine2011}. MHD turbulence is important for a wide range of astrophysical processes, such as the formation and evolution of stars, the transport of energy and momentum, cosmic ray scattering and acceleration \citep{Jokipii1966ApJ,Lerche2001A&A,Qin2002ApJ, YL02, YL_CR08, YLP08, Yan2022rev, Lemoine22,Sampson2023MNRAS} and the amplification of magnetic fields~\citep{Lazarian2020PhPl}. Observational detection of the properties of MHD turbulence in the ISM is essential for a comprehensive understanding of the latter and the processes that take place within it.

The 3D inclination of the magnetic field in interstellar media and its relation to other physical processes is one of the most important scientific questions in the astrophysical community. However, determining the properties of the magnetic field, in particular, its interplay with the ubiquitous interstellar turbulence, is notoriously difficult. Measurement of magnetic field properties mainly relies on two popular observational techniques: polarimetry from synchrotron radiation or dust emission/absorption that only gives the line-integrated or plane of sky magnetic field direction \citep{Davis1951ApJ, Hildebrand2002apsp,2007MNRAS.378..910L,2015ARA&A..53..501A, Hensley2019ApJ}, and Zeeman splitting that gives line-of-sight magnetic field strength in dense clouds \citep{1999ApJ...520..706C,2010ApJ...714.1398C}. Recent effort based on atomic alignment in the magnetic field suggests that the 3D magnetic field topology in the diffuse medium could possibly be measured \citep{YanL06,Kuhn2007ApJ,YanL07,2008ApJ...677.1401Y,2012JQSRT.113.1409Y}, but currently restricted to metal absorption lines due to instrumental restrictions \citep{2020ApJ...902L...7Z}. Therefore, the search for the 3D magnetic field and its underlying relation to turbulence is in a deadlock.

Similar problems happen also for the determination of the nature of MHD turbulence, e.g. sonic and Alfvenic Mach number or MHD mode fractions, in observational data. Several efforts have been made in retrieving the essential parameters in ISM turbulence observations through  statistical techniques. One common method is to measure the power spectrum of electron density fluctuations in the ISM \citep{1995ApJ...443..209A,2010ApJ...710..853C}, which can reveal the presence of turbulent motions. Other methods include studying the velocity statistics in the ISM using spectroscopic observations \citep{2004ApJ...603..180L}, and mapping the distribution of magnetic fields using polarization or spectroscopic measurements (e.g. \citealt{2008ApJ...680..420H}). Despite that, the measurement in earlier studies is either based on crude models of turbulence statistics or qualitative relations between different turbulence observables.

Recent theoretical developments on magnetized turbulence theory suggest that the properties of the magnetic field are encoded in the statistics of MHD turbulence \citep{YL2004ApJ, LP12,2020PhRvX..10c1021M}. Conceptually, MHD turbulence can primarily be decomposed into three modes: Alfv\'en mode, the fast and slow magnetosonic modes (also known as magneto-acoustic modes)~\citep{CL03}. Magnetic field lines are stretched differently by Alfven and magnetosonic modes, and therefore the statistics of magnetic field observables are different. Utilizing this fact, \cite{2020NatAs...4.1001Z} and later \cite{leakage}  suggest that the statistics of polarized synchrotron radiation reflect the fluctuations in the embedded magnetic fields caused by turbulence, which in turn allows us to study the magnetic field and its turbulent properties. In this paper, we aim to develop a new statistical recipe to retrieve the magnetic field inclination angle and the properties of underlying MHD turbulence using simulated datacubes.

 The structure of the paper is as follows. In \S \ref{sec:technique}, we discuss our theoretical construction in measuring the line of sight angle from the theory of turbulence statistics. The numerical simulation setup and its observables are described in \S \ref{sec:method}. Our detailed analysis and decomposed and total cubes results can be found in \S \ref{sec:result}. We discuss the impact of our results in \S \ref{sec:dis}, and we conclude our paper at \S \ref{sec:con}.

\begin{figure*}
\centering
    \includegraphics[width=0.80\textwidth]{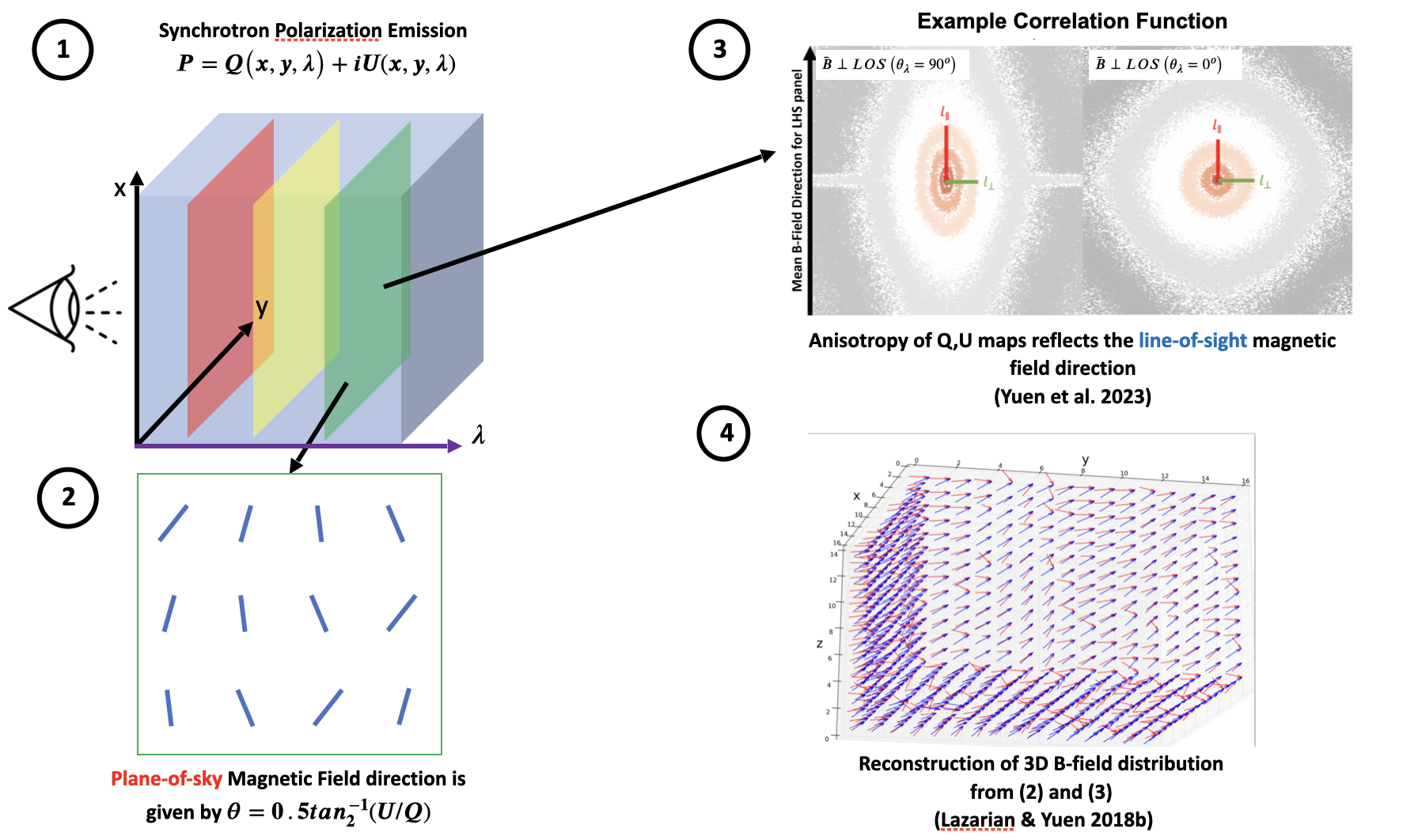}
    \caption{ A figure showing how polarized synchrotron emissions store the information of 3D magnetic field. Panel (1): Emissions from synchrotron emission are stored in a spectral-spatial 3D cube $P=P(x,y,\lambda)$, where the 3rd axis is in the unit of wavelength, along the line of sight. (2). For each $\lambda$, the magnetic field direction is stored in the synchrotron polarization angle
    (3) The anisotropy of Stokes parameters stores the information of the magnetic field inclination. (4) With all this information combined, the 3D magnetic field angle can therefore be reconstructed with appropriate mathematical procedures.}
    \label{fig:method}
\end{figure*}

\section{The essence of the Y-parameter analysis}
\label{sec:technique}

\subsection{Mapping theory of turbulence statistics and their discrepancies}

Let us first briefly summarize the ``Y-parameter" analysis proposed in \cite{leakage}. The fundamental question that statistical theory of MHD turbulence based on the axis-symmetric assumption (\citealt{Monin1971, Shivamoggi1999AnPhy, LP00}, \citealt{YL02}, \citealt{ HaugenPhysRevE, MullerPRL2005, Galtier2009ApJ, BanerjeePhysRevE, VDA}) wants to resolve is ``how observational statistics are mapped from the 3D turbulence statistics". These series of works usually assume a given statistics of 3D turbulence variables (i.e., 3D density ($\rho$), 3D turbulence velocity (${\bf v}$) and magnetic field (${\bf B}$)) in the form of spectral slopes \citep{1995ApJ...443..209A,2010ApJ...714.1398C}, anisotropy measure \citep{2002PhRvL..88x5001C,CL03,MullerPRL2005,2005ApJ...631..320E} and tensor structures \citep{YL02,BanerjeePhysRevE,Verdini2015ApJ,KLP16,KLP17a,2020NatAs...4.1001Z}. The mapping of the 3D statistics to observable statistics is highly nontrivial in interferometry \citep{LP00, VDA} and polarimetry \citep{LP12, LP16}. Notably, both line of sight angles and energy fraction of MHD modes are stored non-linearly in observables' statistics (i.e., spectrum, anisotropy, and tensors).

Attempts have been made to retrieve the line of sight angles and mode fraction from observational data. For instance, earlier attempts asserted that the line of sight angles could be estimated by inspection of polarization percentages~\citep{Clark2018ApJ}. However, these methods are subjected to strong nonlinear interference from ``3D $\rightarrow$ observable" projection, and therefore the accuracy of these methods is questionable. In parallel, the {\it qualitative} analysis of mode fraction has been proposed by \cite{2020NatAs...4.1001Z}, yet there are currently no ways to retrieve the actual quantitative fraction of MHD modes in observations.

\subsection{Y-parameter Science}

Recently, based on the ``mapping theory" of MHD turbulence statistics \citep{LP12,LP16}, \cite{leakage} proposed a method to retrieve both line of sight angle $\theta_{\lambda}$ and mode fraction concurrently via inspection of observable statistics, namely ``Y-parameter analysis". Suppose $X$ is a 2D observable, then $D_{X}$ represents the global correlation function (CF) of the observable $X$:
\begin{equation}
    D_X({\bf R}) = \langle (X({\bf R}') X({\bf R}'+{\bf R}))^2\rangle_{{\bf R}'}
    \label{eq:sf}
\end{equation}
where one can always write $D_X({\bf R})=D_X(R,\phi)$ via series of multipoles (c.f. \citealt{KLP16}):
\begin{equation}
    D_X(R,\phi) = \sum_{m=0,2,4,...}^{\infty} D_m(R) \cos(m\phi)
\end{equation}
where the odd terms of $m$ are zero due to $D_X$ being even. The mapping theory \citep{LP00,LP12,LP16} assumes that the quadrupole-to-monopole ratio ($D_4/D_0$) and higher order terms are small, which as a result $\theta_{\lambda}$ at most quadratic in the statistics of $D_X$. This assumption was used in the previous analyses of anisotropy-related methods (e.g. \citealt{ch5}). However, the appendix of \cite{leakage} showed that in the case of small Alfv\'enic Mach Number $M_A$, none of  $D_{m\ge4}/D_0$ is vanishing, raising concerns about whether studying the statistics of $D_2$ is sufficient in describing the full fluctuations of observables. \cite{leakage} pointed out that the two-point statistics of observables originating from the same turbulence region carry the same spectrum and anisotropy factors. For instance, Stokes parameters in the case of synchrotron emissions are mostly coming from combinations of magnetic fields, whose statistics are derived from the complete functional forms of spectral, anisotropy, and tensor functions (c.f.~\citealt{YL02, leakage}). Distinct Stokes parameters typically possess identical spectral indices and anisotropy scaling; however, their tensor functions, which depend on the line-of-sight angle, have varying forms. Consequently, the observed statistics of Stokes Q and U, for instance, exhibit differences. Since we usually consider second-order statistics, the tensor functions are at most second order (see, e.g. \citealt{LP00}). As a result, the fraction of two-point statistics from two observables will be at most quadratic of $\theta_{\lambda}$.     

Based on the principles mentioned above, \cite{leakage}  suggests that the following parameter
\begin{equation}
Y = \frac{\text{Anisotropy}(D_{I+Q})}{\text{Anisotropy}(D_{I-Q})} = \frac{v/h(D_{I+Q})}{v/h(D_{I-Q})} 
\label{eq:y_central}
\end{equation}
is a measure of line of sight angle $\theta_{\lambda}$ and mode fraction. Where $v$ and $h$ indicate the length in the vertical and horizontal directions with respect to the plane of sky B-field. This Y-parameter is used in the remainder paper.    


\section{Method}
\label{sec:method}
\subsection{Numerical simulations}

In this paper, we employed two publicly available MHD codes ZEUS-MP/3D \citep{2006ApJS..165..188H} and Athena++ \citep{2020ApJS..249....4S} in simulating magnetized turbulence with different physical conditions. We run our simulations for at least two sound crossing times ($\tau_s=L_{box}/c_s$). Our data cubes are time series of three-dimensional, triply periodic, isothermal MHD simulations with continuous force driving via Ornstein–Uhlenbeck forcing, where the smoothing is controlled by $t_{corr}=0.01\tau_{s}$. The energy injection rate is adjusted to simulate various Alfv\'enic Mach numbers, $ M_A$, and plasma $\beta$. The injection is performed so that we only have eddies injected with scales $L_{inj}/L_{box}\ge 1/2$, which corresponds to $0\le|{\bf k}|\le 2$. The driving force contains both incompressible and compressible driving controlled by a free parameter $\zeta$ \citep[see][]{2020PhRvX..10c1021M}:
\begin{equation}
{\bf f} = {\bf f}_{incomp} \zeta + {\bf f}_{comp} (1-\zeta)
\end{equation}
where $\nabla \cdot {\bf f}_{incomp} = 0$. A summary of the simulation parameters is given in Table
~\ref{tab:sim2}. In our calculations, all physical parameters are set to unity unless specified. Notice that isothermal simulations are scale-free, and so units are not an issue in our calculation.

\begin{table*}
\begin{center}
\begin{tabular}{cccccccc}
\hline
\hline
Model Name & Code  & \makecell{Sonic \\Mach \\Number} & \makecell{Alfvenic\\ Mach\\ Number} & \makecell{Plasma\\Beta} & \makecell{Energy\\ Injection\\ Rate} &  Resolution & \makecell{C-Mode\\ Energy\\ Fraction}\\
& & $M_s$ & $M_A$ & $\beta$ &$\epsilon$    & $N_x$ & $f_C$ \\ 
\hline\hline 
AS7 & Athena++& 1.47 & 0.30 & 0.08 & 1.0   & 512  & 0.18 \\
Z15 & ZEUS-MP& 2.13 & 0.52 & 0.12 & 0.1     & 512 & 0.24 \\
Z12 & ZEUS-MP& 1.06 & 0.26 & 0.12 & 0.01    & 512 & 0.24 \\
AS5 & Athena++& 1.43 & 0.29 & 0.08 & 1.0   & 512  & 0.27 \\
AS6 & Athena++& 0.29 & 0.13 & 0.40 & 0.1   & 512 & 0.30  \\
AS3 & Athena++& 1.40 & 0.29 & 0.08 & 1.0    & 512 & 0.32 \\
AS4 & Athena++& 0.28 & 0.13 & 0.45 & 0.1  & 512 & 0.33  \\
Z8 & ZEUS-MP& 0.44 & 0.45 & 2.06 & 0.001    & 512 & 0.34  \\
Z9 & ZEUS-MP& 0.45 & 0.11 & 0.12 & 0.001    & 512 & 0.35  \\
AS1 & Athena++& 1.33 & 0.27 & 0.08 & 1.0    & 512 & 0.35 \\
Z11 & ZEUS-MP& 0.93 & 0.99 & 2.28 & 0.01    & 512 & 0.36  \\
Z5 & ZEUS-MP& 0.15 & 0.15 & 2.00 & 0.0001   & 512 & 0.39 \\
AS2 & Athena++& 0.26 & 0.12 & 0.42 & 0.1    & 512 & 0.41\\
Z4 & ZEUS-MP& 0.15 & 0.61 & 32.88 & 0.0001  & 512 & 0.44  \\
A7 & Athena++& 0.18 & 0.10 & 0.61 & 0.1   & 512  & 0.44 \\
A9 & Athena++& 0.18 & 0.10 & 6.47 & 0.1    & 512  & 0.45 \\
Z6 & ZEUS-MP& 0.14 & 0.04 & 0.13 & 0.0001   & 512 & 0.49 \\
A8 & Athena++& 0.03 & 0.04 & 3.55 & 0.01    & 512 & 0.50 \\
A10 & Athena++&0.03 & 0.04 & 3.55 & 0.01   & 512 & 0.51  \\
A5 & Athena++& 0.17 & 0.10 & 0.69 & 0.1   & 512  & 0.58 \\
A6 & Athena++& 0.03 & 0.04 & 3.55 & 0.01   & 512 & 0.59  \\
A3 & Athena++& 0.15 & 0.09 & 0.72 & 0.1    & 512 & 0.66 \\
A4 & Athena++& 0.02 & 0.04 & 8.00 & 0.01  & 512 & 0.75 \\
A1 & Athena++& 0.13 & 0.09 & 0.95 & 0.1    & 512 & 0.84 \\
A2 & Athena++& 0.02 & 0.03 & 4.50 & 0.01    & 512 & 0.91\\

\hline\hline
\end{tabular}
\end{center}
\caption{\label{tab:sim2} This table contains all the properties of numerical simulations used in the analysis, sorted according to the Compressible mode (C-Mode) Energy Fraction. The cubes with `Z' and `AS' model names are simulated with Zeus and Athena++ solenoidal forcing, whereas the datacubes with `A' model names are generated with Athena++ compressible forcing.}
\end{table*}
\begin{figure*}
\centering
    \includegraphics[width=1.0\textwidth]{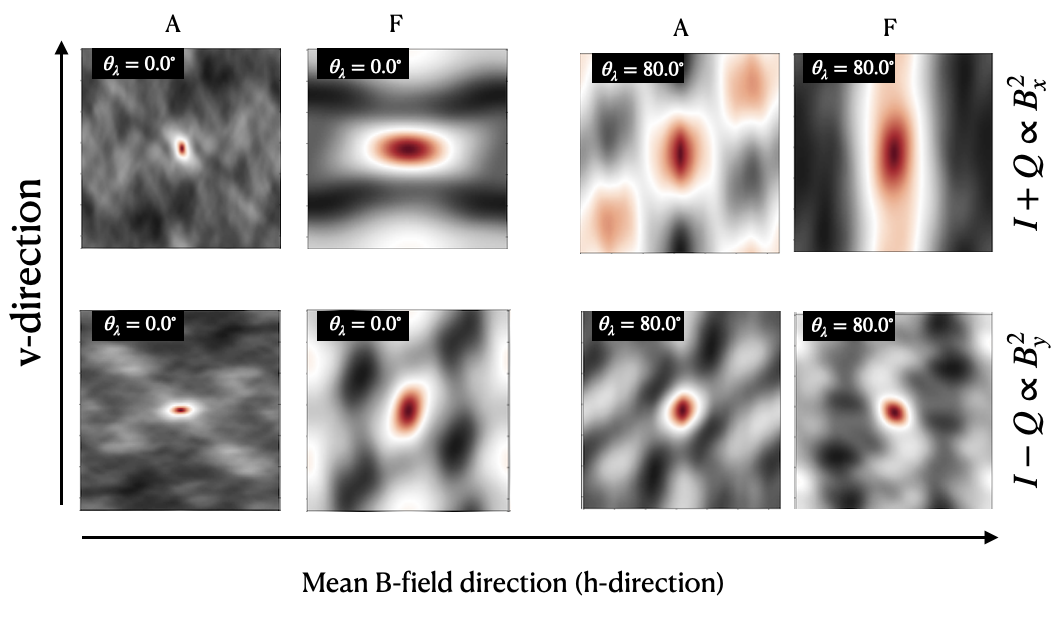}
    \includegraphics[width=1.0\textwidth]{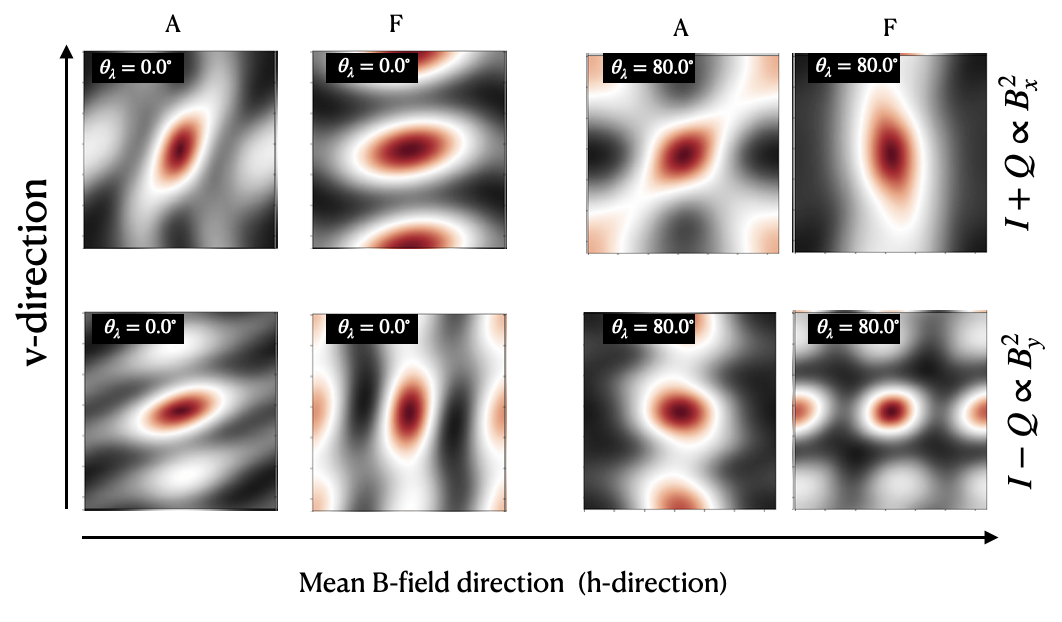}
    \caption{ {\it Upper panel:} The figure shows the correlation function of the decomposed plane of sky components of \textbf{B} field ($B_{x}$, $B_{y}$) of the A (Alfven) and  F (compressible) modes for the solenoidal forcing Z12 MHD data cube with $M_A\sim 0.26$ and $\beta \sim 0.12$  at two $\theta_{\lambda}$ values. The upper and lower sub-panels are for $B_{x}^2$ and $B_{y}^2$, respectively. {\it Lower panel:} same as an upper panel but for the compressible forcing Athena++ A1 cube with $M_A\sim 0.09$. The mode decomposition is performed using the PCA decomposition method described in ~\citep{LP12}. }
    \label{decomposed_anisotropy}
\end{figure*}

\subsection{Synthesis of synchrotron polarization observables}
In general, the synchrotron emission depends on the distribution of relativistic electrons as
\begin{equation}
N_e({\cal E})d{\cal E}\sim {\cal E}^{\alpha} d{\cal E},
\end{equation}
 with the intensity of the synchrotron emission being
\begin{equation}
I_{sync}({\bf X}) \propto \int dz B_{\perp}^\eta({\bf x})
\end{equation}
where ${\bf X} = (x,y)$ is the 2D position of the sky (POS) vector and $B_{\perp} = \sqrt{B_x^2 + B_y^2} $ being the magnitude of the magnetic field perpendicular to the line of sight in $z$-direction. In general, $\eta=0.5(\alpha+1)$ is
a fractional power law. The statistics of $I(\alpha)$ are similar to that of $I(\alpha=3)$ \citep{LP12,2020NatAs...4.1001Z}. Therefore, it suffices to discuss the statistical properties of the case $\alpha=3$.

Per \cite{LP12}, synchrotron complex polarization function {\it with Faraday rotation} is given by:
\begin{equation}
    P_{synch}({\bf R}) = \int dz \epsilon_{synch} \rho_{rel}B^2e^{2i\left(\theta({\bf R},z)+C\lambda^2\Phi(R,z)\right)}
\end{equation}
where $\epsilon_{synch}$ is the emissivity of synchrotron radiation,
\begin{equation}
    \Phi(R,z) = \int_\infty^z dz' (4\pi)^{-1/2}\rho_{thermal}({\bf R},z) B_z({\bf R},z) {\rm rad~m^{-2}}
    \label{eq:chap1.frm}
\end{equation} is the Faraday Rotation Measure \footnote{It is usually more convenient to use $H_z=B_z/\sqrt{4\pi}$ for analysis. }. Notice that $\rho_{rel}$ is the relativistic electron density, while $\rho_{thermal}$ is the thermal electron density. The C-factor $\approx 0.81$  ~\citep{LYLC, malik2020ApJ}. The projected magnetic field orientation is then given by:
\begin{equation}
    \theta_B = \frac{\pi}{2} + \frac{1}{2} \tan^{-1}_2(\frac{U}{Q})
    \label{eq:chap1.Bangle}
\end{equation}
where $\tan^{-1}_2$ is the 2-argument arc-tangent function.  
Also, we will consider only the statistics of $\eta =2$ (i.e. $\alpha = 3$).

\subsection{Mode fraction analysis}
\label{pca}
The MHD mode wave-vectors in the case of isothermal MHD are given by \citep{CL03}:
\begin{equation}
    \begin{aligned}
    \zeta_A(\hat{\bf k},\hat{\lambda}) &\propto \hat{\bf k} \times \hat{\bf  \lambda}\\
    \zeta_S(\hat{\bf k},\hat{\lambda}) &\propto (-1 +\alpha-\sqrt{D}) ({\bf k}\cdot \hat{\bf \lambda}) \hat{\bf \lambda}  \\&+ (1+\alpha - \sqrt{D}) ( \hat{\bf \lambda} \times ({\bf k}\times \hat{\bf \lambda})) \\
    \zeta_F(\hat{\bf k},\hat{\lambda}) &\propto (-1 +\alpha+\sqrt{D}) ({\bf k}\cdot \hat{\bf \lambda}) \hat{\bf \lambda}  \\&+ (1+\alpha + \sqrt{D}) ( \hat{\bf \lambda} \times ({\bf k}\times \hat{\bf \lambda})) \\
    \end{aligned}
    \label{eq:cho}
\end{equation}
where $\alpha = \beta\Gamma/2$, $D=(1+\alpha)^2- 4\alpha\cos ^2\theta_\lambda$, $\cos\theta_\lambda = \hat{\bf k}\cdot \hat{\bf \lambda}$, plasma $\beta\equiv P_{gas}/P_{mag}$ measures the compressibility and $\Gamma =\partial P/\partial \rho$ is the polytropic index of the adiabatic equation of state. The presence of $\hat{\bf k}$ suggests that the direction of the three mode vectors are changing as ${\bf k}$ changes. In this scenario, the perturbed quantities, say for the velocity fluctuations ${\bf v}_1 = {\bf v}-\langle {\bf v}\rangle$ can be written as:
\begin{equation}
    {\bf v}_1({\bf r}) = \int d^3 {\bf k} e^{i{\bf k}\cdot {\bf r}} \sum_{X\in A,S,F} F_{0,X}({\bf k})F_{1,X}({\bf k},\hat{\lambda})  C_X \zeta_X(\hat{\bf k},\hat{\lambda})
    \label{eq:2}
\end{equation}
where $C_X$ denotes the relative weight of the modes and $F_0$, $F_1$ represents the power spectrum, and anisotropy weight, respectively (see \citealt{leakage} and refs therein). The ``magnetic field frame", which we will not cover in this paper, is given by an additional rotation of $\tan\theta_{\lambda}$ from the P(otential)-C(ompressible)-A(lfven) frame $(\hat{\zeta}_P=\hat{\bf k},\hat{\zeta}_A=\hat{\bf k}\times \hat{\lambda}, \hat{\zeta}_C=\hat{\bf k}\times (\hat{\bf k}\times \hat{\lambda}))$. The PCA frame has its special advantage since the ${\bf k}$ sampling is usually complete in $d\Omega_k$. Therefore, we can fix ${\bf k}$ despite other unit vectors changing. For $\zeta_x$, We can always write the arbitrary vector in the Fourier space as :
\begin{equation}
    \zeta_i({\bf k}) = C_P \hat{k}_i + C_C \frac{(\hat{\bf k}\times (\hat{\bf k} \times \hat{\bf  \lambda}))_i}{ |\hat{\bf k} \times \hat{\bf  \lambda}|} + C_A \frac{(\hat{\bf k} \times \hat{\bf  \lambda})_i}{|\hat{\bf k} \times \hat{\bf  \lambda}|}
\end{equation}
which we will name the unit vector $\zeta_{P,C,A}$.

\begin{figure*}
\centering
    \includegraphics[scale=0.8]{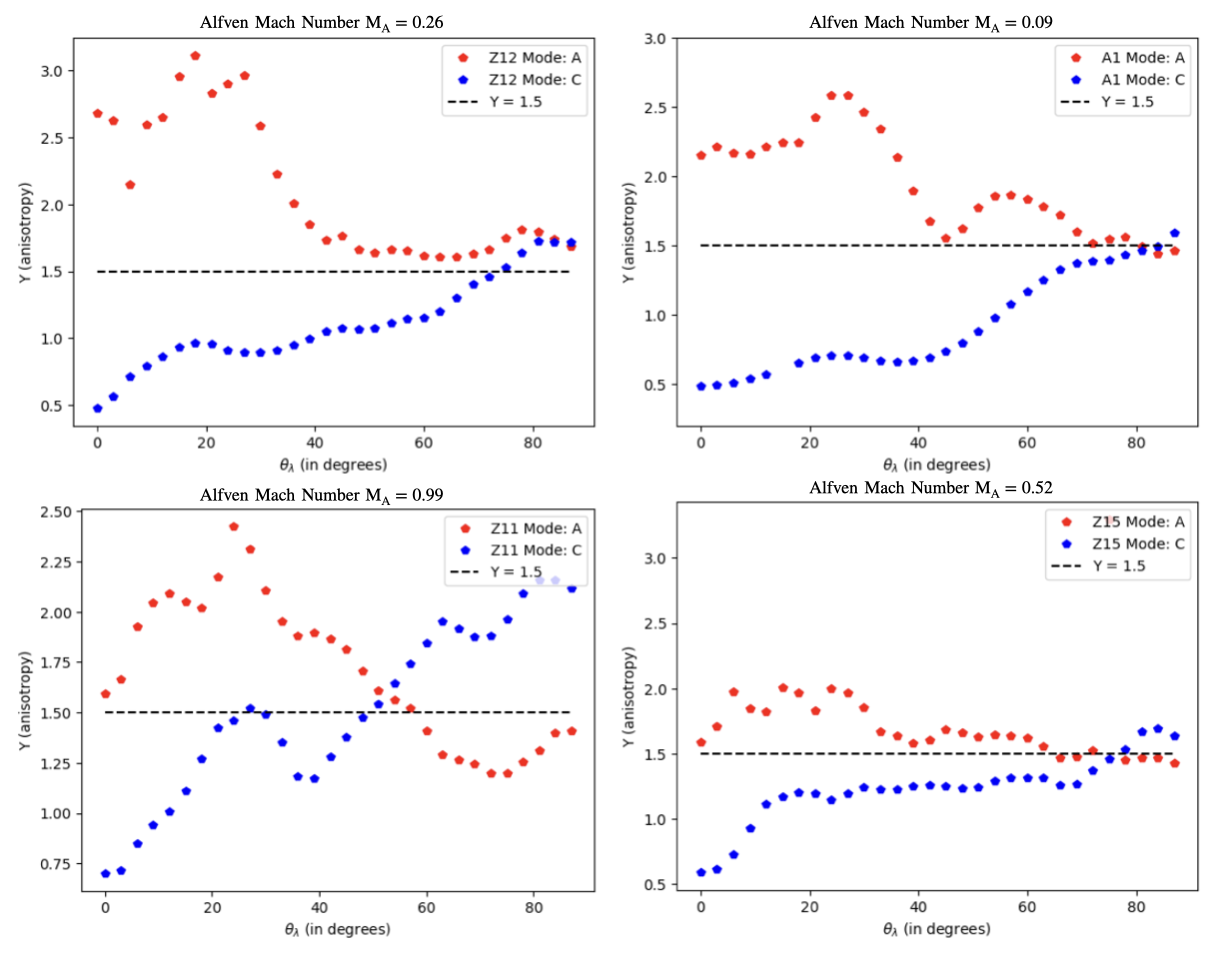} 
    \caption{The plots of Y-parameter vs $\theta_{\lambda}$ for the decomposed \textbf{B} field: pure A (Alfven mode) and pure C (Compressible mode). The red and blue diamond symbol represents the Alfven and Compressible MHD turbulence modes. Where the black dashed line indicates the ${\rm Y} \sim 1.5$. We have plotted four panels for datacube with model names Z12, A1, Z11, and Z15 having different turbulence properties. Here datacube A1 is generated with compressible forcing and is predominantely composed of compressible modes.}
    \label{Ydecomposed}
\end{figure*}

\begin{figure}
\centering
    \includegraphics[scale=0.6]{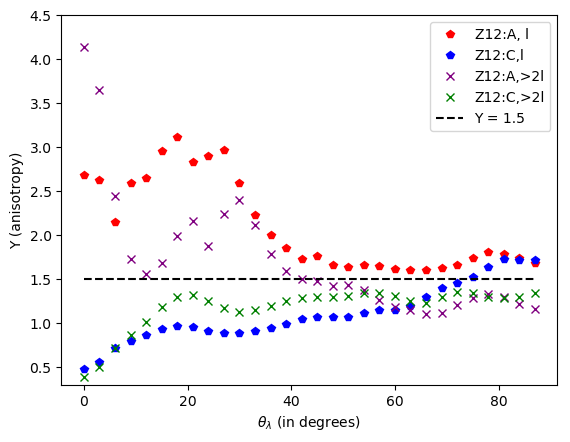}
    \caption{The plot illustrates the variation of the Y-parameter versus $\theta_\lambda$ with length scale. Here, we have computed the Y-parameter for the decomposed B-field at a specific length scale, denoted as `$l$' (indicated by the red and blue diamond symbols for A and C modes). This length scale lies within the injection length scale. Further, we have also evaluated the Y-parameter as a function of $\theta_\lambda$ using a larger length scale of `$2l$' (represented by the purple and green cross symbols for A and C modes), which extends beyond the injection length scale.}
    
    \label{length1}
\end{figure} 

\subsection{Rotation algorithm}
In this paper, we adopt the Euler rotation algorithm\footnote{\url{https://www.github.com/doraemonho/LazRotationDev}} that is adopted in \cite{dispersion} \& \cite{ch5} for obtaining numerical cubes with mean magnetic field at different inclination angle, $\theta_\lambda$. The rotation algorithm in the case of triply periodic numerical cubes has issues in double-counting the corners, which creates artificial effects, especially on the mean of Stokes parameters near $\theta_{\lambda} \approx \pi/4$. To mitigate the effect, one can introduce a spherical filter positioned at the center of the cube. The spherical filter ensures that the basic statistics of Stokes parameters stay regular. For our study, we emphasize that the application of spherical filter does not change the scale-dependent anisotropy of the Stokes parameters correlation function. The rotation matrices are defined as :
\begin{equation}
    \begin{aligned}
    {\bf \hat{T}}_x &= \left[\begin{array}{ccc}
        1 & 0 & 0 \\
        0 & \cos(\theta_x) & -\sin(\theta_x) \\
        0 & \sin(\theta_x) &  \cos(\theta_x) \\
    \end{array}\right]\\
    {\bf \hat{T}}_y &= \left[\begin{array}{ccc}
        \cos(\theta_y) & 0 & \sin(\theta_y) \\
        0 & 1 & 0 \\
        -\sin(\theta_y) & 0  &  \cos(\theta_y) \\
    \end{array}\right]\\
    {\bf \hat{T}}_z &= \left[\begin{array}{ccc}
        \cos(\theta_z) & -\sin(\theta_z) & 0 \\
        \sin(\theta_z) &  \cos(\theta_z) & 0 \\
        0 & 0&  1\\
    \end{array}\right]\\
    \end{aligned}
    \label{eq:euler_rotation}
\end{equation}
where we will write the rotation matrix ${\bf \hat{T}} = {\bf \hat{T}}_x{\bf \hat{T}}_y{\bf \hat{T}}_z $ and $\theta_{x,y,z}$ are desired rotations along the x,y,z axis respectively.

For rotations of 3D scalar cubes, say $\rho({\bf r})$ with ${\bf r}\in \mathcal{R}^3$ , the new cube is given by $\rho({\bf \hat{T}}^{-1}{\bf r})$. For vector cubes ${\bf x}({\bf r})$, the new cube is given by ${\bf \hat{T}}{\bf x}({\bf \hat{T}}^{-1}{\bf r})$. The inverse transform is invoked for the position vector ${\bf r}$ because the rotation of cubes is equivalent to rotating the observing frame in the opposite direction.

\section{Analysis and Results}
\label{sec:result}
 
 To establish the Y-parameter as a reliable technique, we test this method to numerical MHD cubes with various plasma properties, as outlined in Table~\ref{tab:sim2}. Although ISM compositions can be intricate, we aim to simplify our analysis by focusing on a combination of independent emitting layers and not considering the impact of Faraday rotations, as shown in the first panel of Figure~\ref{fig:method}. These emitting layers contain fluctuating small-scale magnetic fields, shown in panel 2 of Figure~\ref{fig:method}. In addition to these small-scale components, our data cubes exhibit a mean magnetic field since $M_A\lesssim 1$. By analyzing these data cubes, we can gain insight into the nature of MHD turbulence and orientation of B-field in the ISM plasma and further develop our understanding of this important astrophysical phenomenon.

\subsection{Analysis of decomposed data cubes}

In order to gain a better understanding of the behaviour of the Y-parameter for various MHD turbulent plasma modes and its dependence on $\theta_{\lambda}$, we initially decomposed the magnetic fields of our MHD cubes into two general components: Alfven mode (A) and compressible modes (F or C, which we use interchangeably), utilizing the PCA decomposition techniques as briefly described in \S~\ref{pca} (see  \citealt{LP12} for more details). We then proceeded to investigate the anisotropies of projected observables $I+Q \propto B_{x}^2$ and $I-Q \propto B_{y}^2$ for each of these modes separately. It is worth noting that due to their different dependence on the $B_{\perp}$ and $B_{\parallel}$ components, the anisotropy also inherently depends on $\theta_{\lambda}$. To illustrate the variation of anisotropy, we have presented two cases in Figure~\ref{decomposed_anisotropy}, one with $\theta_{\lambda} \sim 0.0\degree$ and another at $\theta_{\lambda} \sim 80.0\degree$. The upper panel of Fig.~\ref{decomposed_anisotropy} shows the orientation of the correlation functions for both A and F decomposed modes for the solenoidal-forcing Z12 MHD datacube. Meanwhile, the lower panel of Figure~\ref{decomposed_anisotropy} shows the correlation function orientations for the compressible-forcing A1 cube. 

To calculate the Y-parameter using the correlation function distributions, we adopted a fitting approach in the central region with Gaussian functions along both directions. The presence of a smooth distribution of correlation function contours on the fitting scale is a prerequisite the analysis. By analyzing these figures, we can draw the following conclusions:

\begin{enumerate}
    \item Low $\theta_{\lambda}$ case: The anisotropy of $B_{x}^2$ and $B_{y}^2$ for the A mode is perpendicular and parallel to the projected mean magnetic field, respectively. However, it is vice versa for the case of F mode. 

    \item High $\theta_{\lambda}$ case: These anisotropies of $B_{x}^2$ and $B_{y}^2$ are more or less the same irrespective of A and F mode.  
    \item The relative anisotropies (Eq.~\ref{eq:y_central}) for $B_{x}^2$ and $B_{y}^2$  can give us a parameter that can be used to characterize the MHD turbulence modes in the given turbulent media. Due to its dependency on $\theta_{\lambda}$, it can be used as a probe to retrieve the mean field inclination angle in ISM.    
\end{enumerate}

To investigate the relationship between the Y-parameter and the mean-field inclination angle, $\theta_{\lambda}$, we analyzed the Y-parameter for decomposed A and F modes obtained from four distinct MHD turbulence datacubes, as illustrated in Fig.\ref{Ydecomposed}. Our analysis revealed that the Y-parameters for the A and F modes exhibit different characteristics. Specifically, the Y-parameter for the pure F (compressible) mode shows an increasing trend as $\theta_{\lambda}$ increases. In contrast, the Y-parameter for the pure A (Alfven) mode exhibits a decreasing trend with $\theta_{\lambda}$. Even though they were generated using different driving mechanisms (the left upper panel of Fig.\ref{Ydecomposed} depicts the Z12 cube with solenoidal forcing, while the right upper panel depicts the A1 cube with compressible forcing), the functional dependency of the Y-parameter for pure F and A modes exhibit similar functional dependencies on $\theta_{\lambda}$. Notably, the average {\it Y-parameter} has clear separation at the value of $\rm{Y} \sim 1.5$, which could be used as a criterion to distinguish between MHD turbulence modes. Additionally, we would like to highlight that the relative anisotropy, Y-parameter, is estimated through the fitting of the Correlation Function, as depicted in Fig. \ref{decomposed_anisotropy}, using length scales that fall within the injection length scale. In Fig. \ref{length1}, we observe that when the length scale is within the inertial range, the decomposed B-field exhibits contrasting trends in the Y-parameter and converges toward $\sim 1.5\pm0.5$ at higher $\theta_\lambda$. However, for tests using larger length scales that lie outside the inertial range, the Y-parameter began to converge towards the statistical demarcation of $\sim 1.5\pm0.5$ even at smaller $\theta_\lambda$.

\begin{figure*}
\centering
    \includegraphics[scale=0.8]{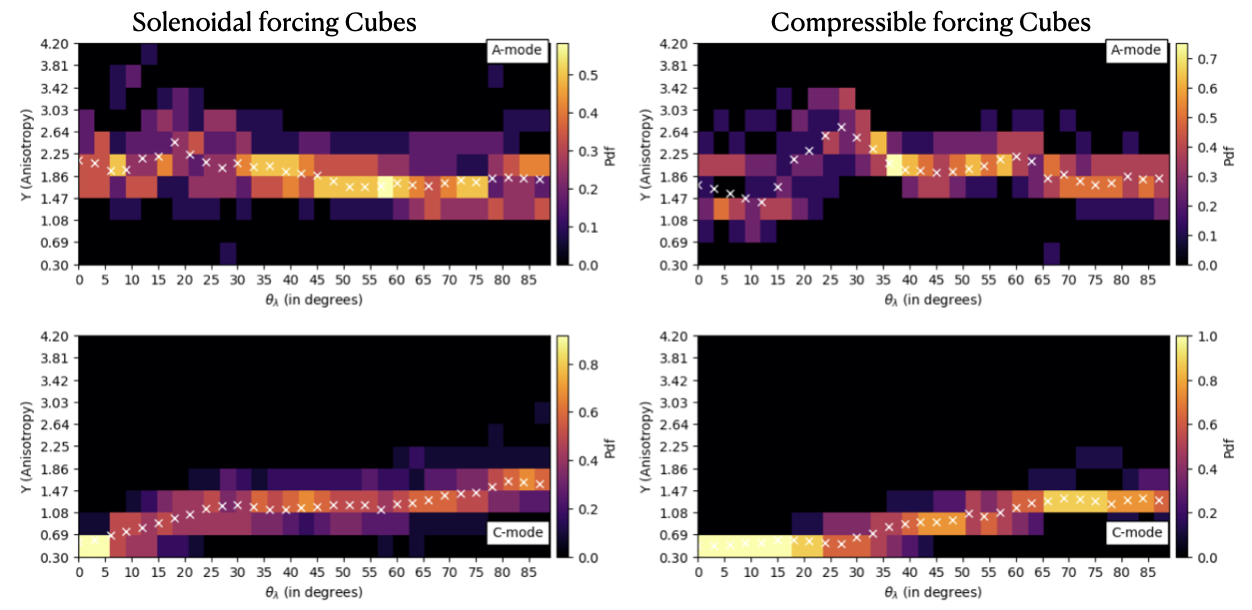}
    \caption{{\it Left Panel:} To enhance the statistics of the Y-parameter for mode-decomposition, in this plot,
    we show the probability density and median Y-parameter for the decomposed \textbf{B} field against the inclination angle of total mean-field, $\theta_{\lambda}$. We have taken the median and variance of the Y-parameter for 12 MHD turbulence cubes with $f_C<0.4$ (see Table~\ref{tab:sim2} for more details). The black color represents the vacancy of data points and hence probability is zero. The color bar represents the PDF at every value of $\theta_{\lambda}$. The upper and lower sub-panels are for purely Alfv\'enic and Compressible modes, respectively. The white cross indicates the median values of the Y-parameter. {\it Right Panel:} It is the same as the left panel except for MHD cubes. Here we have taken 8 MHD turbulence cubes with C-modes dominance.}
    \label{combined_Ydecomposed}
\end{figure*}

To replicate the complex composition of the ISM, we employed the Y-parameter as a mode decomposition technique on 25 datacubes of MHD turbulence (${\rm M_{A}<1.0}$), spanning a wide range of plasma properties, energy injection rates, and driving mechanisms (see Table~\ref{tab:sim2}). While analyzing the individual cubes, we found that the technique gives ambiguous results when the compressible mode energy fraction of the cubes lies in the range of $0.5<f_C<0.6$. Therefore, to ensure the validity of our results, we excluded 5 datacubes with C-mode energy fractions falling between 0.5 and 0.6. The remaining dataset comprised of 12 turbulence cubes subjected to solenoidal forcing, exhibiting less than 40\% of energy in the Compressible turbulence mode, and 8 datacubes characterized by compressible-driven turbulence, with a dominating energy fraction in the compressible mode. We investigated the correlation function of $B_{x}^2$ and $B_{y}^2$ for the individual modes decomposed from all these cubes and found that they follow a similar trend, as shown in Fig.\ref{decomposed_anisotropy}. Furthermore, we evaluated the relative anisotropies, or ``Y-parameter," for each cube and found an equivalent trend for the Z12 and A1 cubes, as demonstrated in the upper two panels of Fig.\ref{Ydecomposed}.

To improve the statistical significance of our results, we calculated the statistical probability density and median of the Y-parameter for the A and C modes in their respective decomposed data cubes at every $\theta_{\lambda}$ value. The probability density (PDF) and median Y-parameter against $\theta_{\lambda}$ is presented in Fig.\ref{combined_Ydecomposed}, where the left panel represents the 12 MHD turbulence cubes with $f_C < 0.4$. In contrast, the right panel comprises 8 compressible dominant datacubes with $f_C>0.5$. The upper and lower sub-panels in Fig.\ref{combined_Ydecomposed} show purely Alfv'enic and compressible modes, respectively. Notably, the functional dependency of the Y-parameter on $\theta_{\lambda}$ in both these plots for the Alfven and compressible modes is consistent with the behaviour of the Y-parameter in the individual data cube. Thus it can act as a powerful diagnostic for the reconstruction of 3D magnetic fields and understand the mode-decomposition of MHD turbulence.

\subsection{Analysis of Total Datacube}

\begin{figure*}
\centering
    \includegraphics[width=1.0\columnwidth]{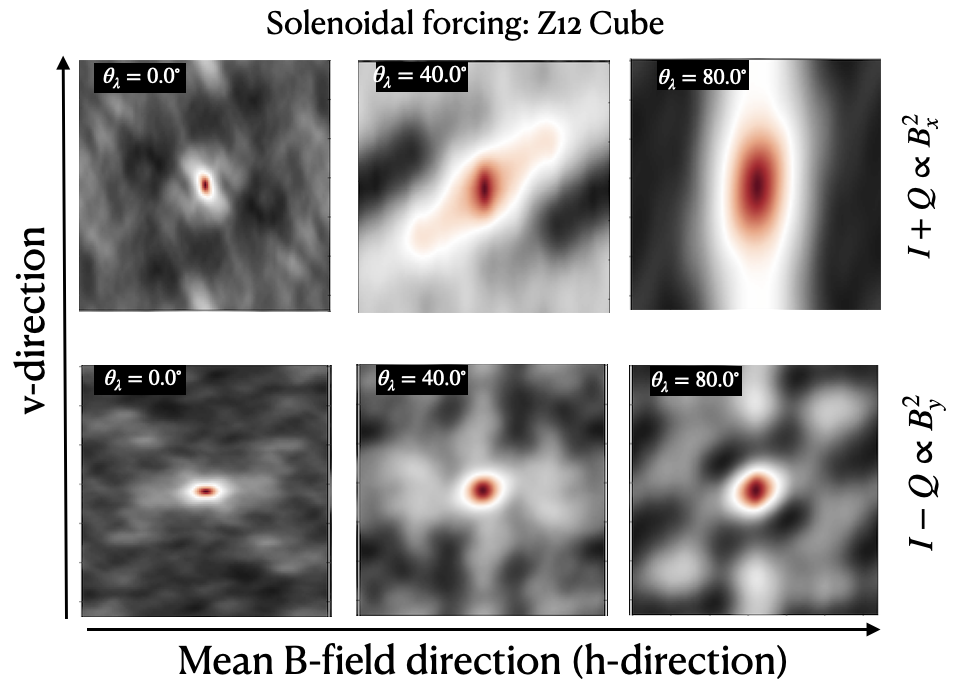}
    \includegraphics[width=1.0\columnwidth]{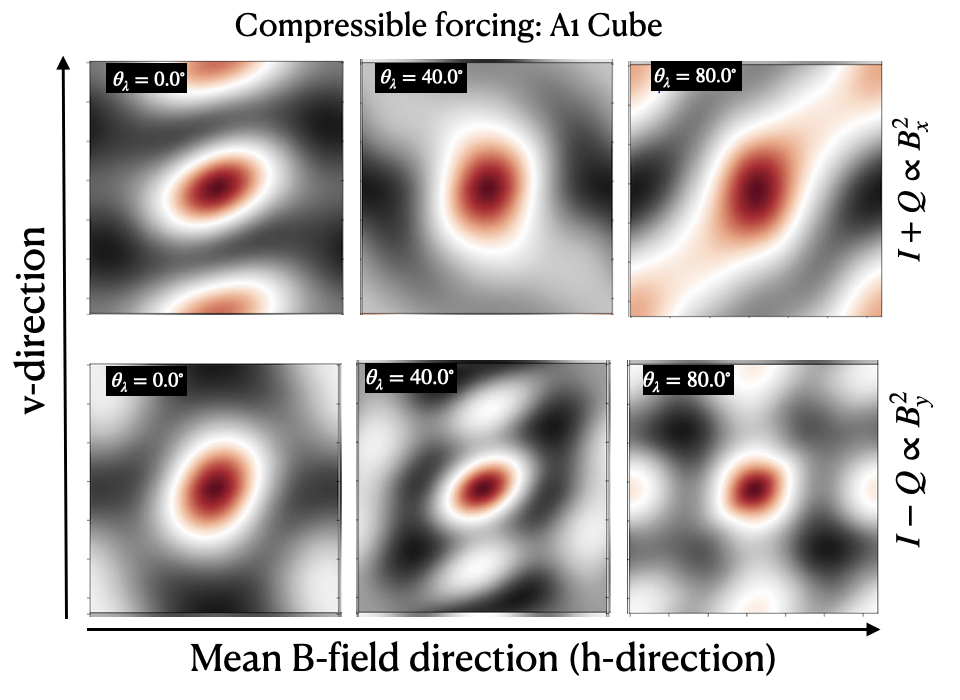}
    \caption{ {\it Left panel:} The figure shows the correlation function of the plane of sky components of the total \textbf{B} field for the solenoidal forcing Z12 MHD data cubes with $M_A\sim 0.26$ and $\beta \sim 0.12$  at $\theta_{\lambda} \sim$ $0.0\degree$, $40.0\degree$, $80.0\degree$. The upper and lower panel is for $B_{x}^2$ and $B_{y}^2$, respectively. {\it Right panel:} This is the same as the left panel but for the compressible-forcing A1 MHD cube.}
    \label{total_z6_A1_anisotropy}
\end{figure*}
 
\begin{figure*}
\centering
    \includegraphics[scale=0.8]{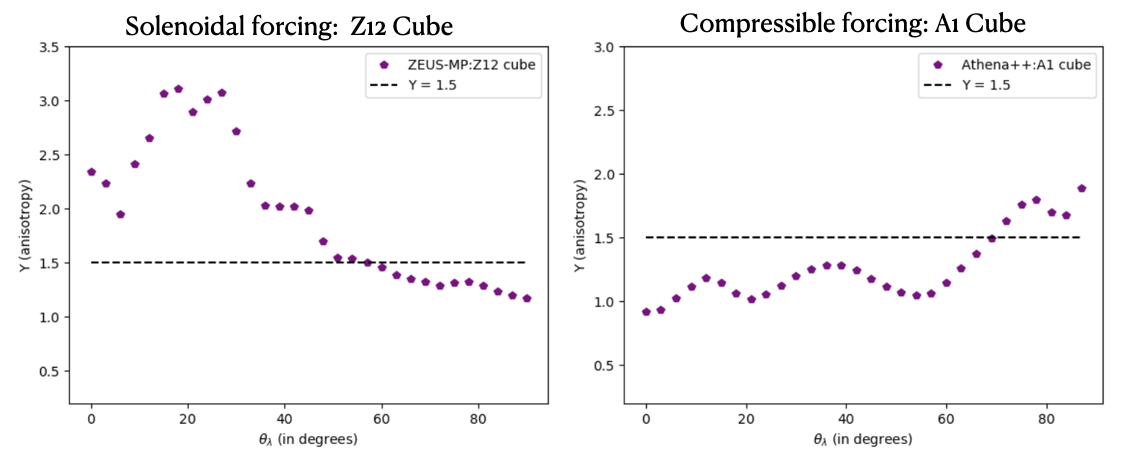}
    \caption{ {\it Left Panel:} The plot of the Y-parameter for the total \textbf{B} field for the Z12 MHD cube with $\theta_{\lambda}$. The black dashed line represents the threshold value of ${\rm Y}=1.5$. {\it Right Panel:} This plot is the same as the left panel but for the A1 MHD cube.}
    \label{singleY_totalB}
\end{figure*}

\begin{figure*}
\centering
    \includegraphics[height=10cm,width=17cm
]{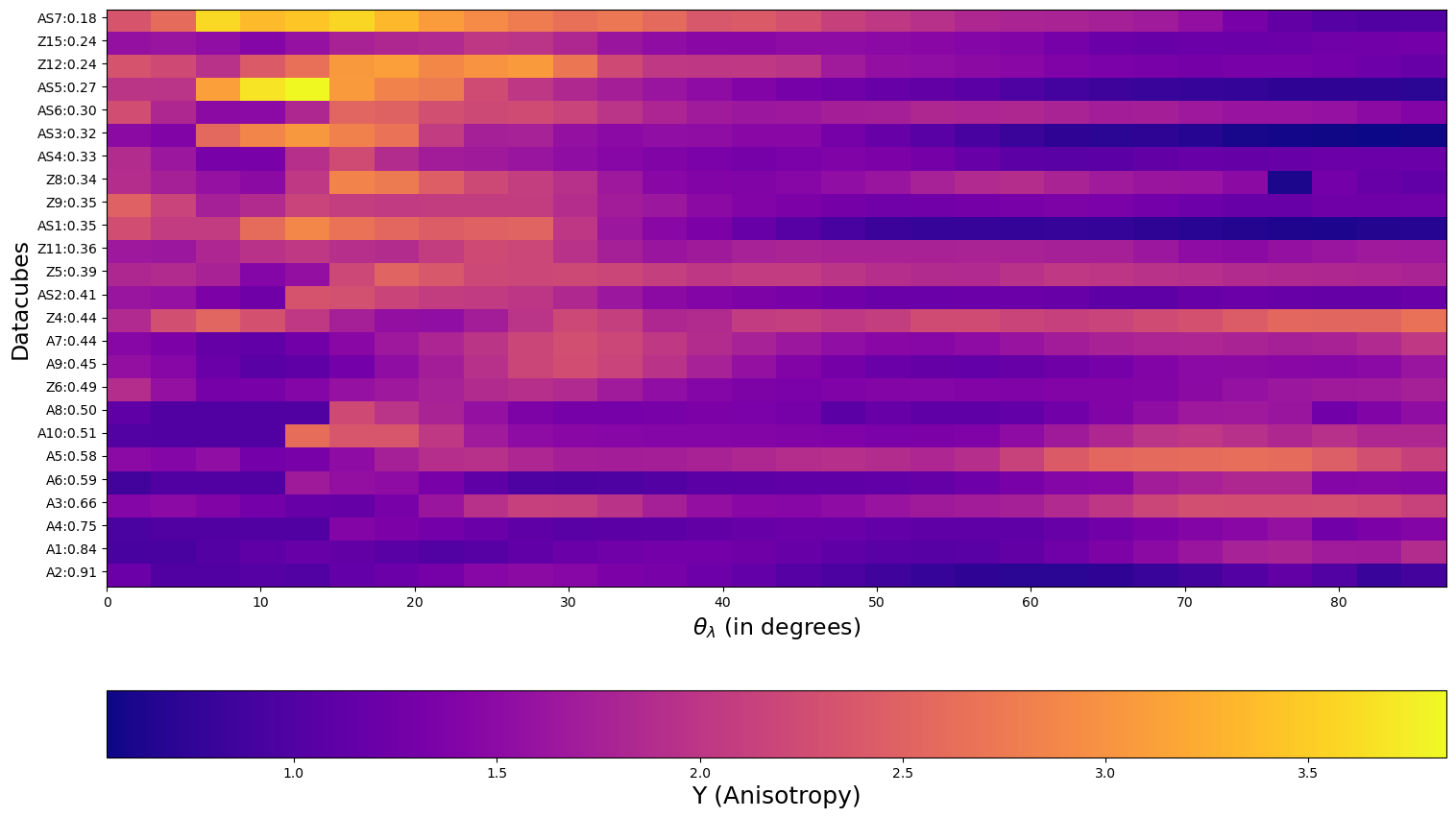}
    \caption{The heatmap shows the distribution of the Y-parameter for 25 MHD datacubes. Here x and y-axises represent the $\theta_{\lambda}$ and different data cubes sorted in descending order of $f_C$, respectively.}
    \label{everyY_totalB}
\end{figure*}

MHD turbulence in the real scenario is always complex and exhibits a mixture of turbulence modes. To validate our technique for the reconstruction of 3D magnetic fields and to understand the MHD turbulence mode, we investigate the total B-field of 25 datacubes listed in Table~\ref{tab:sim2}. We begin by estimating the correlation function of $B_{x}^2$ and $B_{y}^2$ for the total field and examining their behaviour with respect to $\theta_{\lambda}$. The correlation functions of two cubes are plotted in Fig.\ref{total_z6_A1_anisotropy}. The left panel of this figure shows that the correlation functions for the Z12 MHD cube follow the correlation functions distribution for the decomposed Alfven mode in the upper panel of Fig.\ref{decomposed_anisotropy}. Since the Z12 MHD cube is solenoidally driven, the Alfven energy fraction is dominant and therefore it is natural that Alfv\'enic "signature" overwhelms that of the compressible one. Similarly, the correlation function anisotropy for the A1 MHD cube with the total field in the right panel of Fig.\ref{total_z6_A1_anisotropy} is presided over by the correlation function of the pure compressible mode, as seen in the lower panel of Fig.\ref{decomposed_anisotropy}. It motivates us to evaluate the relative anisotropies as a Y-parameter for the total field in these MHD turbulence cubes. We calculated the Y-parameter from the turbulence data cubes with varying inclination angles of the total field and plotted them in Fig.\ref{singleY_totalB}. It is clear from this figure that the Y-parameter for these cubes follows the trend of ${\rm Y} > 1.5$ for Alfven mode dominance (Z12) and ${\rm Y} < 1.5$ for compressible mode dominance (A1). We conducted a thorough analysis of the Y-parameter's behavior by examining all of our MHD cubes. We then created a heatmap of the Y-parameter against $\theta_{\lambda}$ in ascending order of $f_C$, which is shown in Fig.~\ref{everyY_totalB}. We observed that all these cubes maintained their respective trend depending on the mode energy fraction. To avoid ambiguity in the Y-parameter due to approximate similar mode energy fraction, we excluded 5 datacubes from our simulation sets with $f_C$ falling between 0.5 and 0.6. The rest of the total B-field analysis is with the remaining 20 data cubes where 12 datacubes have $f_C<0.4$ and 8 datacubes with $f_C>0.5$ with ${\rm M_{A}<1.0}$.

\begin{figure*}
\centering
    \includegraphics[scale=0.9]{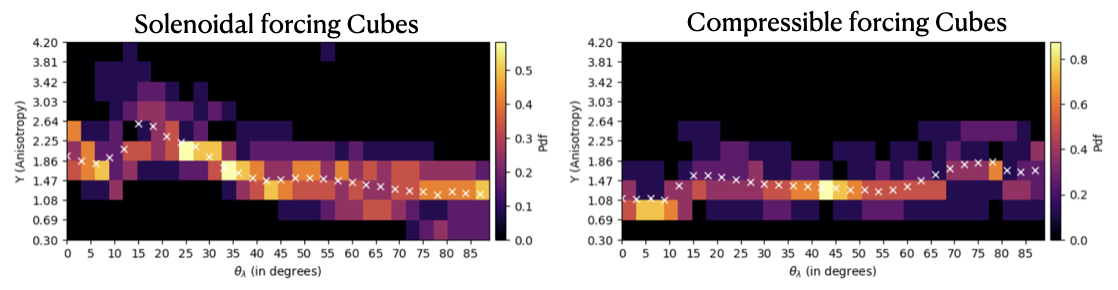}
    \caption{{\it Left panel:} The plot shows the probability density and median Y-parameter for the PoS total \textbf{B} field against the $\theta_{\lambda}$. We have used 12 MHD cubes with $f_C<0.4$ with Alfven modes dominance. The black color represents the vacancy of data points; hence, the probability is zero. The color bar represents the PDF at every value of $\theta_{\lambda}$. The white cross indicates the median values of the Y-parameter. {\it Right Panel:} The plot shows the probability density and median Y-parameter versus $\theta_{\lambda}$ for the 8 MHD turbulence cubes with C-modes dominance ($f_C>0.5$).}
    \label{pdf_totaly}
\end{figure*}

Moreover, to strengthen the statistical confidence of the Y-parameter as a diagnostic for 3D magnetic field and MHD turbulence mode, we evaluated the probability density (PDF) and median of the Y-parameter at every value of $\theta_{\lambda}$. The PDF and median Y-parameter are shown in Fig.~\ref{pdf_totaly}, and they follow the respective decomposed mode trends. It indicates that the Y-parameter can serve as an important tool for understanding the 3D magnetic field structures and underlying MHD turbulence in a region.

\begin{figure*}
\centering
    \includegraphics[height=10cm,width=17cm]{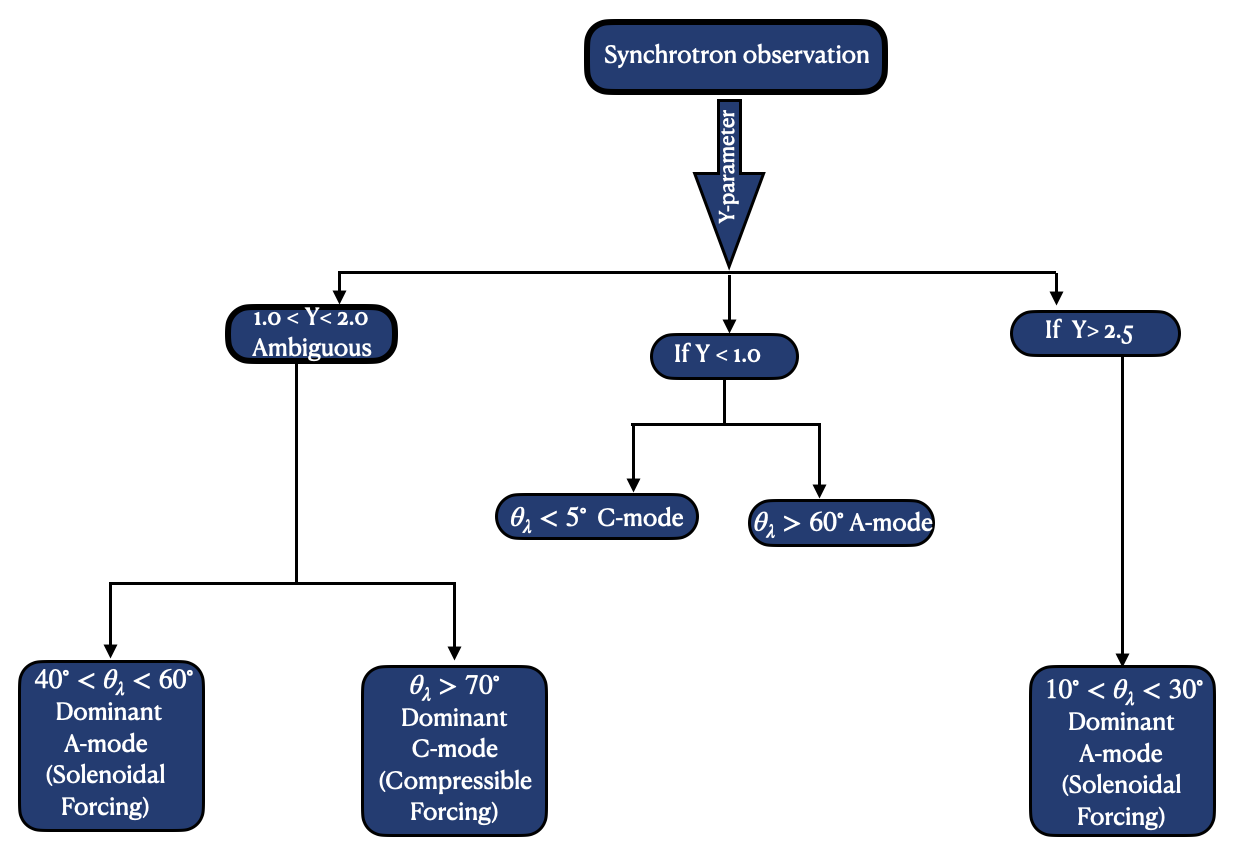}
    \caption{Flowchart to elaborate the scheme for estimating mean field inclination angle and identification of dominant turbulence mode. Here A and C represent the Alfven and Compressible modes.}
    \label{flowchart}
\end{figure*} 

  As a result, from Fig.~\ref{pdf_totaly}, we can conclude the following inferences for $\theta_{\lambda}$: 
        \begin{itemize}
              
            \item If ${\rm Y}\gtrsim 2.5$, we have $10^\circ<\theta_{\lambda}<30^\circ$ and Alfven dominance i.e. $f_C < 0.4$. 
            
            \item If ${\rm Y}\lesssim 1.0$ there are two possibilities: (i) $\theta_{\lambda}$ is $\lesssim 5^\circ$ and turbulence has C modes dominance i.e. $f_C > 0.5$ (ii) $\theta_{\lambda}$ is $\gtrsim 60^\circ$ and turbulence is dominated by A-mode. 
             
            \item  $60\degree \gtrsim \theta_{\lambda} \gtrsim 40\degree$ and turbulence is Alfv\'enic or $\theta_{\lambda}$ is $\gtrsim 70^\circ$ and the turbulence is compressible if Y-parameter is in the intermediate range;
            
        \end{itemize}

We have also elaborated our scheme for estimating the magnetic field inclination angle and identifying the dominant MHD turbulence mode in the flowchart shown in Fig~\ref{flowchart}. It is important to note that the medium's rotation measure influences this technique's applicability. In the case of MHD turbulence simulated datacubes, we have observed that this technique effectively determines the $\theta_\lambda$ values in regions where the emitting layer Faraday rotation and its variance remains within a range of $50^\circ$.

From the above discussion, it is clear that the observed statistics of MHD turbulence modes depend on the region's mean magnetic field orientation. As a result, this can be a crucial diagnostics to investigate the 3D magnetic field in the ISM and extended objects such as SNR and PWN.  

\section{Discussion}
\label{sec:dis}
\subsection{Implication to general studies of ISM magnetic field and turbulence}

Our paper extends the formalism of the two-point studies in synchrotron analysis \citep[see, e.g.][]{LP12}. The relationship between the statistics of various observables and the intrinsic statistics of 3D MHD turbulence variables is a complex and ongoing question within the ISM community \citep{1995ApJ...443..209A,ElmegreenScalo,Boldyrev2006PhRvL,2007ARA&A..45..565M, Draine2011}. It is mainly due to the absence of true sampling of 3D data like the solar wind community (see, e.g. ~\citealt{siqi2021ApJ, siqi2022apj, siqi2023arXiv}) and the complications of line-of-sight projections (See, e.g. the same problem for velocity variables, \citealt{LP00,VDA}). Only in very rare circumstances one can obtain measurements of 3D information in observations, for instance, ground state alignment \citep{Landi2004ASSL,YanL06,Kuhn2007ApJ,YanL07,2008ApJ...677.1401Y,2012JQSRT.113.1409Y,2020ApJ...902L...7Z}, dust estimation \citep{Davis1951ApJ,Hildebrand2002apsp,2007MNRAS.378..910L,2015ARA&A..53..501A, Hensley2019ApJ,2020A&A...633A..51Z} or starlight polarization from GAIA\footnote{https://gea.esac.esa.int/archive/}. These measurements, unfortunately, are sparsely measured to date, which needs to be revised for analysis of the properties of interstellar media (See some attempts in ISM like \citealt{2023MNRAS.518..919S}). Techniques that allow retrieval of intrinsic turbulence properties in ISM is therefore of great importance (e.g. VCS, \cite{LP00}).

More recently, the ISM community has developed a few ways of obtaining the turbulence parameters in synchrotron observations, namely sonic Mach number (\citealt{2011Natur.478..214G,2012ApJ...755L..19B}), Alfvenic Mach number \citep{dispersion}, turbulence index \citep{LP16,2023MNRAS.518..919S} and mode signatures \citep{2020NatAs...4.1001Z}. The remaining puzzles regarding the search of turbulence properties are (i) the relative orientation of the magnetic field angle to the line of sight; (ii) {the dependence of energy modes composition diagnosis on the magnetic field inclination angle}. Based on the qualitative study of \cite{leakage}, in this paper, we have presented a statistical technique to recover the LOS inclination angle $\theta_{\lambda}$ of the magnetic field with various turbulence parameters, which is important for studies of magnetic field geometry and relevant sciences that depend on the magnetic field orientation.

\subsection{Application to ongoing and upcoming radio survey} 

Radio surveys such as the Low-Frequency Array (LOFAR)\footnote{https://lofar-surveys.org/index.html}, and Square Kilometer Array (SKA)\footnote{https://www.skao.int/en} are currently underway or planned for the near future, promising high-resolution and high-sensitivity radio observations of the interstellar medium (ISM) and galaxies~\citep{Gitti2018A&A}. However, current diagnostic techniques are limited to Faraday rotation and polarization measurements of synchrotron emission \citep{LOFAR2019A&A, Basu2019Galax, Heald2020Galax, LOFAR2023MNRAS}, which are sensitive to different components of magnetic fields and are affected by the line-integrated effect. Therefore, there is a pressing need to develop new diagnostic techniques. Our work proposes a well-suited technique for diffuse radio observations from these surveys ~\citep{LOFAR2019A&A, LOFAR2023MNRAS}. By analyzing the relative anisotropies involved in observable Stoke parameters, our technique can determine magnetic field orientation and complement the estimation of magnetic field geometry using Faraday rotation, leading to more accurate estimates of the strength and the 3D inclination of the B-field in the ISM. 

It is important to note that this technique is designed for the length scales within the inertial ranges. In the case of ISM and any extended sources such as PWN and SNRs, the injection length scales of the turbulence can vary from $50-150$ pc, thus the required anisotropy for the Y-parameter analysis can be readily captured at scales $5-10$ pc according to the scale separations in our synthetic observations. The coherence length should be first identified from observations using the well-known structure/correlation function. One such application is discussed briefly in \ref{monogem}. Additionally, our analysis enables systematic investigation of the composition of MHD turbulence modes in different plasma environments.

\subsection{Implication to observations of Pulsar TeV's halo} 
\label{monogem}
 To ensure the reliability of our analysis to investigate the mean field inclination angle and turbulence mode classification, we have applied this technique to archival radio polarisation observations from the $5\degree \times5\degree$ region of Pulsar Wind Nebulae. The radio observations were conducted using the Effelsberg 100-m telescope at a wavelength of $21 {\rm cm}$ with the resolution of $9.5$ arcmin \citep{Uyaniker1999}. Our detailed analysis of this region will be presented in an upcoming publication \citep{malik2023}.

In brief, this region encompasses the Monogem Pulsar, also known as pulsar B0656+14. Recent observations by the High-Altitude Water Cherenkov Observatory (HAWC) have revealed a spherical TeV halo around both the Monogem and Geminga pulsars~\citep{Abeysekara2017Sci}. To explain the origin of such high-energy emission, ~\cite{Liu2019PRL} proposed a model involving an anisotropic diffusion model to explain the TeV observations from the Geminga source. One of the crucial parameters for their analysis is the mean magnetic field inclination angle of the region.

We applied our analysis to the radio polarisation observations to estimate the mean field inclination angle for the Monogem region and found that the Y-parameter is below the 1.5 range. It suggests that the region has primarily {\it compressible} turbulence, with a low mean-field inclination angle. These outcomes are consistent with theoretical predictions for such a halo from  
~\cite{Liu2019PRL}. 

\subsection{Impact of our work on the studies of magnetic field strength via DCF}

Obtaining the value of $\theta_{\lambda}$ also allows us to determine the magnetic field strength more accurately. The recent debate about the validity of the \citeauthor{1951PhRv...81..890D}-\citeauthor{1953ApJ...118..113C} technique (\citeyear{1951PhRv...81..890D,1953ApJ...118..113C}) suggests that the DCF technique is inaccurate in estimating magnetic field strength, with errors up to $1000$ times \citep{2021A&A...647A.186S}. The call for revision of the DCF technique is urgent as most B-field strength estimation in ISM and molecular cloud studies are done by the DCF technique (see, e.g. \citealt{2022FrASS...9.3556L} for a review).  The correction of the DCF technique requires extra turbulence information from interstellar media: Alfv\'enic Mach number, mode fraction, and line of sight angle, spatial co-existence of dust and spectral line origins \citep{ch5, PYC23}. Our work complements the series of works on correcting the DCF technique in obtaining the true B-field strength in the sky.

\section{Conclusion}
\label{sec:con}

In summary, we have developed a novel diagnostic technique to investigate three-dimensional magnetic fields and underlying MHD turbulence in the ISM. Our technique is primarily based on polarisation observations of the ISM and represents a significant advancement in the field, as no similar diagnostic exists to date. To validate the efficacy of our technique, we tested it extensively using simulated MHD turbulence data cubes with varying plasma properties, energy injection rates, and driving mechanisms (solenoidal or compressible forcing). These simulations also included differences in the composition of Alfven and Compressible modes. Our results show that our technique can successfully retrieve the mean-field inclination angle, $\theta_{\lambda}$, along the line of sight and the underlying dominant MHD turbulence mode. Our major findings are listed below;

\begin{itemize}

    \item If ${\rm Y}\gtrsim 2.5$, we have  $10^\circ<\theta_{\lambda}<30^\circ$ and Alfven dominance i.e. $f_C < 0.4$.
    
    \item If ${\rm Y}\lesssim 1.0$ there are two possibilities: (i) $\theta_{\lambda}$ is $\lesssim 5^\circ$ and turbulence is compressible (i.e. $f_C > 0.5 $) (ii) $\theta_{\lambda}$ is $\gtrsim 60^\circ$ and turbulence has A-mode dominance. 
    
    \item  $60\degree \gtrsim \theta_{\lambda} \gtrsim 40\degree$ and turbulence is Alfv\'enic or $\theta_{\lambda}$ is $\gtrsim 70^\circ$, and the turbulence is compressible if Y-parameter is in the intermediate range $1.5\pm0.5$;
    \item We found a statistical demarcation of ${\rm Y} \sim 1.5\pm0.5$ (with ${\rm Y} > 1.5$ for A mode and ${\rm Y} < 1.5$ for C-mode) to obtain the dominant fraction of MHD turbulence modes.
    
\end{itemize}

Hence, our method offers a notable progression in interstellar research, permitting a more detailed exploration of 3D magnetic fields and MHD turbulence. It effectively reveals valuable insights into ISM turbulence.



\section*{Acknowledgments} 
We thank the anonymous referee for valuable comments that significantly improved the paper. SM would like to thank P. Pavaskar and SQ Zhao for the helpful discussions. SM and HY gratefully acknowledge the computing time granted by the Resource Allocation Board and provided on the supercomputer Lise and Emmy at NHR@ZIB and NHR@Göttingen as part of the NHR infrastructure. The calculations for this research were conducted with computing resources under the project bbp00062. The research presented in this article was supported by the Laboratory Directed Research and Development program of Los Alamos National Laboratory under project number(s) 20220700PRD1. This research used resources provided by the Los Alamos National Laboratory Institutional Computing Program, which is supported by the U.S. Department of Energy National Nuclear Security Administration under Contract No. 89233218CNA000001. This research also used resources of the National Energy Research Scientific Computing Center (NERSC), a U.S. Department of Energy Office of Science User Facility located at Lawrence Berkeley National Laboratory, operated under Contract No. DE-AC02-05CH11231 using NERSC award FES-ERCAP-m4239 (PI: KHY, LANL).

\section*{Data Availability}
The MHD turbulence datacubes simulated in this paper are available upon reasonable request from the corresponding author.

\bibliographystyle{mnras}
\bibliography{refs}
%
%
\bsp	
\label{lastpage}
\end{document}